\newcommand{\etal}{et al.\ }
\newcommand{\lya}{Ly$\alpha$\ }

\newcommand{\nh}{N_{\rm HI}} 
\newcommand{\kms}{\;{\rm km}\,{\rm s}^{-1}}

\newcommand\cdunits{{\rm cm}^{-2}}
\newcommand\ggh{{\Gamma_{\rm HI} }}
\newcommand{\hmpc}{h^{-1}{\rm Mpc}}

\documentstyle[11pt,AATS,epsf]{article}	

\begin{document}	

\title{The Evolution of the Lyman Alpha Forest From $z\sim 3\rightarrow 0$}

\author{Romeel Dav\'e}
\affil{Steward Observatory, Univ. of Arizona, 933 N. Cherry Ave.,
Tucson, AZ, USA}


\begin{abstract}

I review results obtained from studies of the high-redshift \lya forest,
and present new results from HST/STIS spectra of low-redshift quasars in
comparison with cosmological hydrodynamic simulations.  The evolution
of the \lya forest from $z\sim 3\rightarrow 0$ is well-described by
current structure formation models, in which \lya forest absorbing gas
at all redshifts traces moderate-overdensity large-scale structures.
I describe some of the insights provided by hydrodynamic simulations
into the observed statistical trends of \lya absorbers and the physical
state of the absorbing gas.

\end{abstract}


\section{Introduction} 

Quasar absorption lines have been used as probes of the high-redshift
universe for some time (e.g. Sargent et al. 1980), since the \lya
(1216\AA) transition is redshifted into the optical at $z\ga 2$.
HIRES (Vogt et al. 1994) on the Keck 10m telescope has been at the
forefront in providing high S/N, high resolution quasar
spectra that has enabled dramatic advances in our understanding of the
high-redshift intergalactic medium (Rauch 1998).  Conversely, studies
of low-redshift \lya absorbers require ultraviolet spectroscopy,
which has only recently become routinely possible with the {\it
Hubble Space Telescope}.  The HST Quasar Aborption Line Key Project
(Bahcall et al. 1993) observed over 80 quasars using the Faint Object
Spectrograph (FOS), and found a surprisingly large number of absorbers as
compared to an extrapolation from high-redshift (Bahcall et al. 1996;
Weymann et al. 1998).  Recently the deployment of the Space Telescope
Imaging Spectrograph (STIS) has provided a dramatic increase in
resolving power (resolution $7\kms$ as opposed to FOS's $230\kms$),
thereby fully resolving all \lya absorbers seen in UV quasar spectra.
While only the brightest quasars and AGN may be easily observed with STIS,
this has still resulted in significant gains in our understanding of
low-redshift \lya absorbers and their relation to high-redshift absorbers.
In these proceedings I report some recent results from HST/STIS.

In conjunction with these observations, hydrodynamic simulations of
structure formation in currently favored cosmological scenarios have
elucidated a new paradigm for the nature of \lya forest absorbers,
particularly at high redshift.  These simulations indicate that \lya
absorbers arise in highly photoionized diffuse intergalactic gas
tracing the dark matter in non-equilibrium large-scale structures.
Their temperature is set by a balance between photoioinization heating
from a metagalactic UV background (presumably from quasars; Haardt \&
Madau 1996) and adiabatic cooling due to Hubble expansion, resulting
in a tight relation $T\propto \rho^{0.6}$.  These simulations are able
to reproduce various statistical properties of high-redshift \lya
absorbers in detail, spanning the range of damped \lya systems to the
weakest observed detected with HIRES, all within a model that generically
arises in CDM cosmologies (Hernquist et al. 1996; Dav\'e et al. 1997).
Figure~1 shows a comparison of a portion of the HIRES spectrum of
Q1422+231 (Songaila \& Cowie 1996) with a simulated spectrum covering
the same redshift interval, having resolution and noise properties that
emulate the data.  Can you tell which is which? (answer at the end).

Due to the relatively simple relationship between \ion{H}{1} and the
underlying dark matter, a formula known as the ``Fluctuating Gunn-Peterson
Approximation" (FGPA; see Croft et al. 1998 for full formula) provides
a remarkably accurate description of \lya absorbing gas:
\begin{equation}
\tau_{HI} \propto \rho^{1.6} \ggh^{-1},
\end{equation}
where $\tau_{HI}$ is the \ion{H}{1} optical depth, $\rho$ is the
density of dark matter (or baryons), and $\ggh$ is the
metagalactic \ion{H}{1} photoionization rate incident on the absorber.

\section{Results From High-Redshift \lya Forest Studies}

The FGPA has been combined with HIRES quasar spectra to provide
stringent constraints on many aspects of the physics of high-redshift
intergalactic gas.  For instance, a measurement of the mean optical
depth together with an independent estimate of $\ggh$ result in a 
robust estimate of $\Omega_b\approx 0.02 h^{-2}$ (Rauch et al. 1997),
in good agreement with D/H measurements (Burles \& Tytler 1998)
and recent CMB analyses (e.g. Pryke et al. 2001).  

The FGPA also implies that a given \lya optical depth (or flux, in the
optically-thin regime) corresponds to a particular density of gas at a
particular temperature.  Hence ionization corrections can be accurately
obtained to constrain the metallicity of the diffuse high-redshift
intergalactic medium (IGM) from metal-line observations of \ion{C}{4}
(Songaila \& Cowie 1996; Rauch, Haehnelt \& Steinmetz 1997; Dav\'e et
al. 1998).  These studies found [C/H]$_\odot\approx -2.5$ for a Haardt \&
Madau (1996) ionizing background shape, and significantly lower for a
spectrum with fewer high-energy photons (as would be expected if Helium
had not yet reionized; see Heap et al. 2000).  Furthermore, recent
VLT/UVES observations of \ion{O}{6} absorption suggests the presence of
some metals even in voids (Schaye et al. 2000), presenting a significant
theoretical challenge to models of metal ejection and transport in the
IGM (e.g. Aguirre et al. 2001).

The most ambitious use of the FGPA, and the one for which it was
originally developed, is a reconstruction of the matter power spectrum
at $z\sim 3$ from Keck/HIRES quasar spectra.  Since the optical depth in
the \lya forest traces the matter density, fluctuations in the optical
depth reflect a 1-D probe of the amplitude and shape of the matter power
spectrum, at scales of $\sim 1\rightarrow 10\hmpc$ (comoving).  For a
flat universe, this measurement yields a constraint on the matter density
of $\Omega_m\approx 0.5+0.29(\Gamma-0.15)$, where $\Gamma$ is the power
spectrum shape parameter (Croft et al. 2000).  This value is slightly
higher that in the ``concordance model" ($\Omega_m\approx 1/3$; Bahcall
et al. 1999), but the latest CMB measurements combined with the $H_0$
Key Project value (Mould et al. 2000) suggest a higher $\Omega_m$ as well
(Pryke et al. 2001).  Thus the \lya forest provides an independent avenue
for doing ``precision cosmology", alongside CMB, Type Ia supernovae, etc.

\begin{figure}[hb]	
\plotfiddle{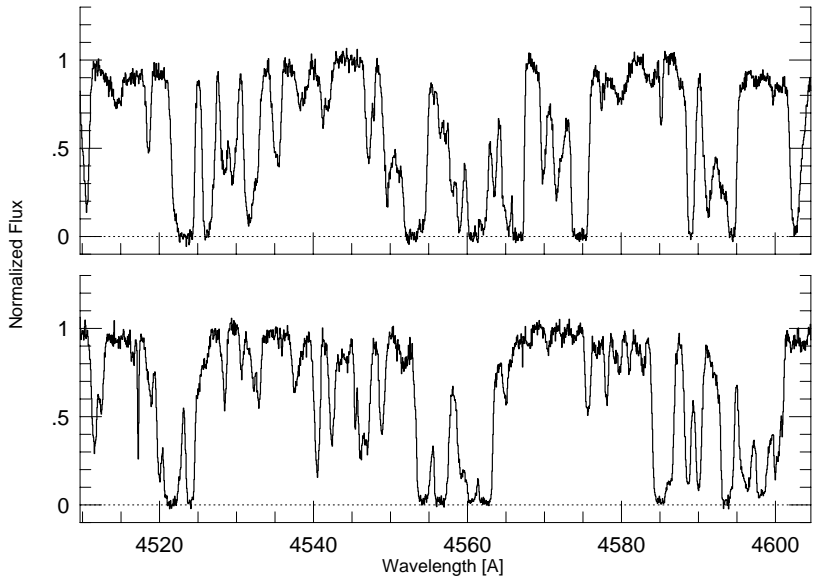}{4.8in}{90}{45}{45}{180}{80}
\caption{100\AA\ section from the spectrum of Q1422+231, and an artificial
spectrum drawn from a hydrodynamic simulation of a $\Lambda$CDM universe,
having resolution and noise characteristics constructed to be similar
to Q1422.  The fluctuating absorption pattern seen in the simulated
spectrum (which one is it?) arises naturally from large-scale structure
generated by hierachical collapse, and is statistically indistinguishable
from that seen in Q1422. }
\end{figure}

\section{Evolution of the IGM to Low Redshift}

Recently, hydrodynamic simulations have been evolved to redshift zero to
investigate the nature of low-redshift \lya absorbers and the evolution of
the \lya forest.  They indicate that while many of the baryons have moved
into galaxies, clusters, and a diffuse shocked ``warm-hot intergalactic
medium" (Cen \& Ostriker 1999; Dav\'e et al. 2001), roughly one-third of
all baryons continue to reside in photoionized gas tracing large-scale
structure.

These simulations are able to explain a wide range of Key Project
observations quite naturally within the context of hierarchical
structure formation scenarios.  For instance, the Key Project data
showed that high-redshift ($z\ga 2$) \lya absorbers are evolving away
quite rapidly, whereas for $z\la 1.5$ the absorber number density
evolution slows abruptly and dramatically (Bahcall et al. 1996).
In the simulations of Dav\'e et al. (1999) and Theuns et al. (1998),
this occurs due to the dimunition of the quasar population providing the
metagalactic photoionizing flux.  As a result, the rapid evolution of
absorbers due to Hubble expansion in the high-$z$ universe is countered
by the increase in neutral fraction of \lya absorbing gas at $z\la 2$.
These simulations disfavor a scenario in which a different population
of absorbers dominates at low redshift as compared to high redshift,
and instead suggest that the nature of absorbing gas is quite similar.

\begin{figure}[hb]	
\plotfiddle{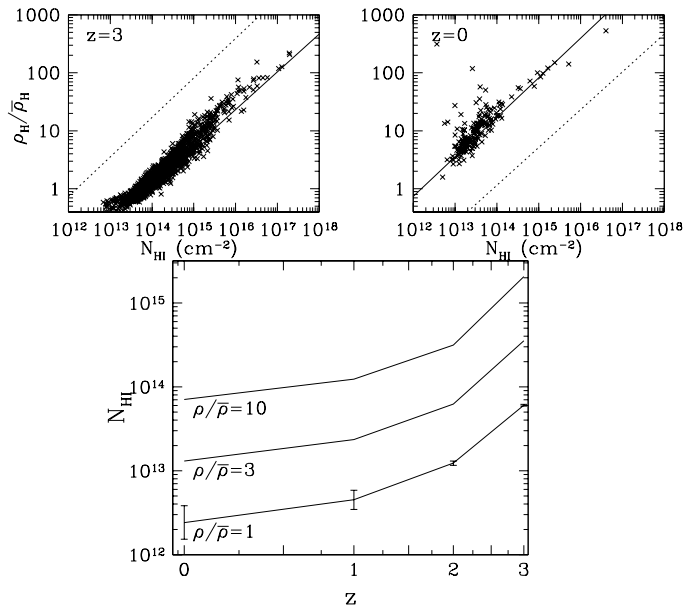}{2.9in}{0}{45}{45}{-120}{-50}
\caption{{\it Top panels:} Relation between column density of absorber
and maximum physical density (in units of mean cosmic density) at the
absorber peak optical depth, at $z=3$ and $z=0$.  Best-fit relation
is shown as the solid line, and is reproduced in the other panel as
the dotted line.  {\it Bottom panel:} For three selected densities,
the evolution of the corresponding absorber column density as a function
of redshift.  The $z\sim 0$ forest is qualitatively quite similar
to the $z\sim 3$ forest, except column densities are
shifted down by $\sim 1.5$ dex.}
\end{figure}

Weymann et al. (1998) found that stronger absorbers disappear more rapidly
than weaker absorbers, and in fact the weakest absorbers seen in the FOS
sample (rest equivalent width $W_r\approx 0.1\AA$) show an {\it increase}
in number between $z\sim 1.5\rightarrow 0$.  Dav\'e et al. suggested
that this arose due to differential Hubble expansion, in particular that
low-density regions expand faster than high-density ones and therefore
provide an increasing fractional cross-section of absorption to lower
redshifts.

Motivated by the good agreement between simulations and FOS observations,
Dav\'e et al. investigated the physical state of the gas giving rise
to \lya absorption at low redshift.  They found that the gas follows
similar physical relationships as deduced at high-redshift, and is thus
of the same basic character, but that a given column density absorber
corresponds to different physical densities at different redshifts.

This effect is shown in Figure~2.  The top left panel shows the
density-column density relation at $z=3$ from a $\Lambda$CDM simulation.
The tight relation reflects the accuracy of the FGPA; much of the
scatter about the best-fit relation (solid line) comes from errors in
Voigt profile fitting (noise with S/N=30 was added to the artificial
spectra to roughly match observations).  At low redshift, the relationship
persists, though there is more scatter due to the presence of absorbers
arising in shock-heated intergalactic gas (Dav\'e et al. 1999; Cen \&
Ostriker 1999).  The best-fit slope of the $\rho-\nh$ relation is similar,
but the amplitude has shifted considerably from $z=3\rightarrow 0$.
This shift arises because of the interplay between Hubble expansion and
the evolution of the ionizing background.

The bottom panel of Figure~2 focusses on the evolution of this
trend with redshift.  Here, we show the typical column density of an
absorber arising in gas with three different densities, as a function of
redshift.  For instance, gas at the mean density (which typically remains
around that density over a Hubble time) produces a strong absorber of
$\nh\approx 10^{14}\cdunits$ atr $z=3$, but by $z=0$ it produces only
a very weak absorber of $\nh\approx 10^{12.5}\cdunits$.  In general,
a given density corresponds to a column density $\sim\times 30$
lower at $z=0$ as compared to $z=3$.  Thus studying physically and
dynamically equivalent absorbers at high and low redshift requires
comparing across different column densities.

Using this trend, the knowledge and intuition gained from studies of
the \lya forest at high redshift can be translated to expectations at
low redshift.  For instance, it is known that most absorbers with $\nh\ga
10^{14.5}\cdunits$ at $z\sim 3$ are enriched (Songaila \& Cowie 1996);
thus one expects that absorbers with $\nh\ga 10^{13}\cdunits$ should be
enriched to at least a similar level (though detecting these metal
lines will be challenging).

\section{Low-Redshift \lya Forest Observed with HST/STIS}

A new window on the low-redshift \lya forest has opened with the
deployment of STIS aboard HST.  STIS's E140M Echelle grating provides
$\sim 7\kms$ resolution, a level of detail comparable to HIRES albeit with
somewhat more noise.  STIS can thus probe weak forest absorbers that are
predicted to be physically similar to those at high-redshift, and allows
a detailed comparison to be made with hydrodynamic simulations.

Dave \& Tripp (2001) compared the quasar spectra of PG0953+415 and
H1821+643 to carefully-constructed artificial spectra drawn from a
$\Lambda$CDM simulation.  The statistical properties of absorbers in the
simulations were in remarkable agreement with observations, as shown in
Figure~3.  The column density distributions agree down to the smallest
observable absorbers, in both slope and amplitude.  Interestingly, the
slope of the column density distribution (from ${d^2 N\over dzd\nh}\propto
\nh^{-\beta}$) is measured to be $\beta\approx 2.0\pm 0.2$, which is
considerably steeper than at high redshift ($\beta\approx 1.5$; Kim et
al. 2000).  This indicates, in agreement with Weymann et al.,
that stronger absorbers have evolved away faster than weaker ones.
Our result is also in broad agreement ($\sim 1\sigma$ higher) with Penton
et al. (2000), who found $\beta\approx 1.8$ from GHRS data.

\begin{figure}[hb]	
\plotfiddle{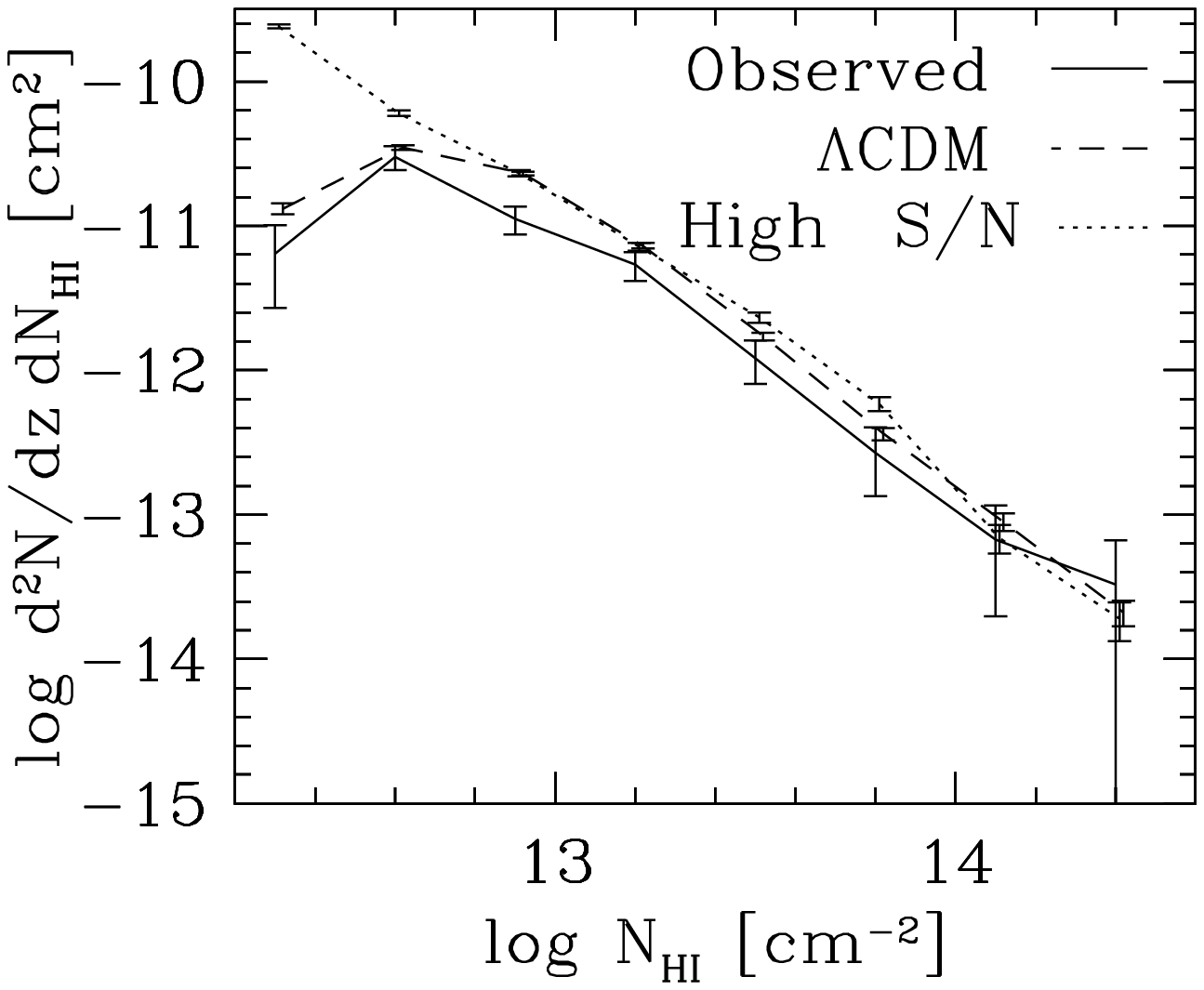}{0cm}{0}{45}{45}{-220}{-310}
\plotfiddle{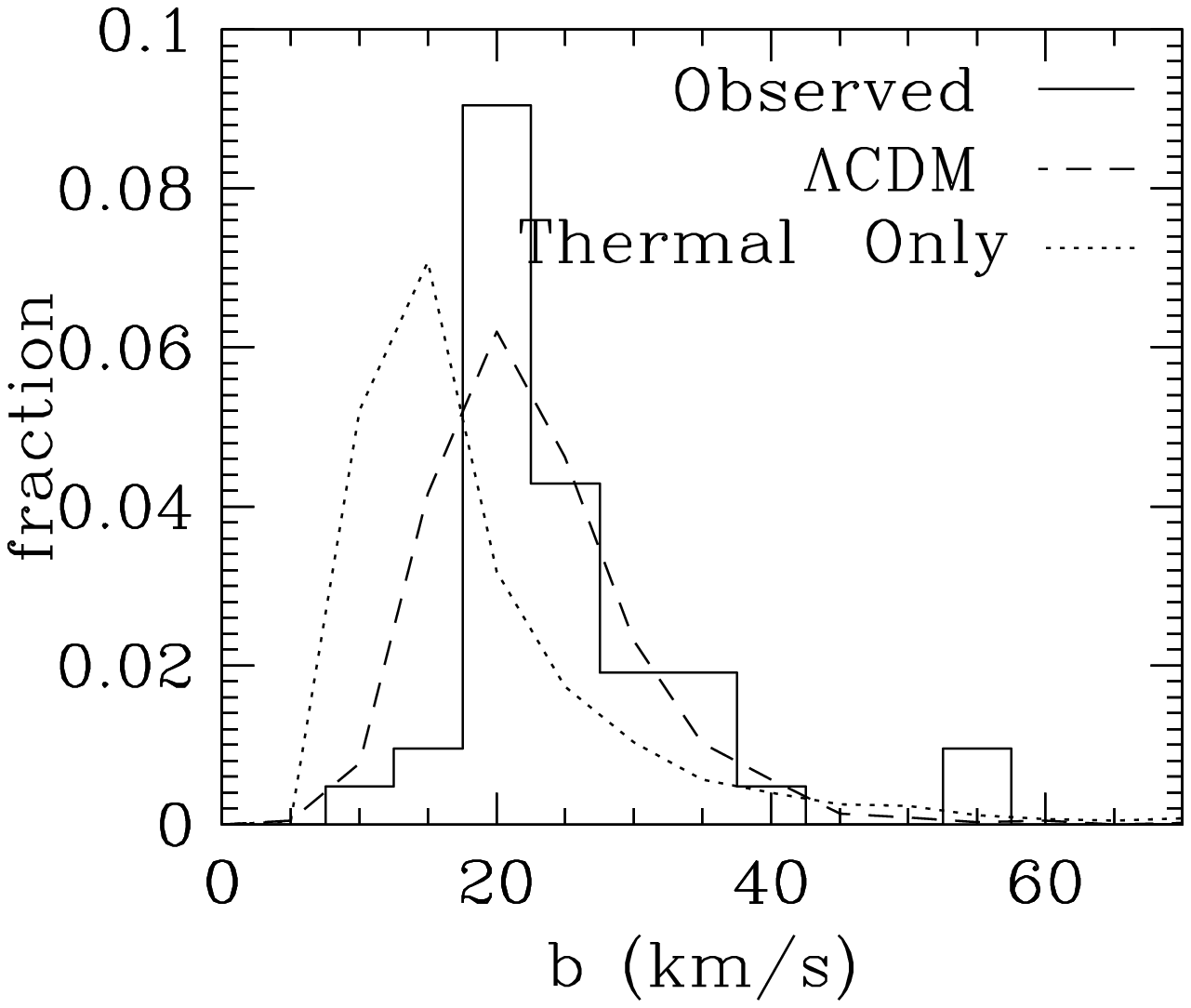}{4cm}{0}{45}{45}{-20}{-185}
\caption{Column density (left panel) and $b$-parameter (right panel)
distributions from HST/STIS observations of PG0953 and H1821 (solid
lines), compared with artificial spectra (dashed lines).  Dotted line
in left panel shows the result of increasing S/N by $\times 5$ in
the artificial spectra, indicating absorber sample is complete down
to $\nh\la 10^{13}\cdunits$.  Dotted line in right panel shows the 
simulated line widths purely from thermal broadening, indicating 
that thermal broadening provides a substantial component of line
widths at low redshift. }
\end{figure}

The distribution of line widths ($b$-parameters) is also in excellent
agreement with observations.  The median $b$-parameter is around $21\kms$,
but this value is sensitive to the column density detection threshold,
because the column density and $b$-parameter are correlated.  This can
be understood from the fact that higher column density systems arise in
higher density gas that has higher temperature.  Since thermal broadening
contributes significantly to line widths (compare the dotted and dashed
lines in the right panel of Figure~3), higher column density absorbers
will tend to have larger widths.  Thus Shull et al. (2000), who find a
median $b$-parameter of $\approx 30\kms$ from FUSE and GHRS data, are
not in disagreement with our result because their sample has a higher
column density detection limit.

The large thermal component of linewidths facilitates a measurement
of the typical temperature of \lya absorbing gas at low redshift.
By comparing the observed median $b$-parameter to simulations, Dav\'e \&
Tripp found that purely photoionized \lya absorbers (those that have not
been shock-heated) at the mean density have a temperature of $T\approx
5000$K, in agreement with Ricotti, Gnedin \& Shull (2000).

The agreement in amplitude of the column density distribution can be
translated into a constraint on the \ion{H}{1} photoionization rate
$\ggh$, since varying $\ggh$ would shift the simulated column density
distribution proportionally in the horizontal direction (assuming
optically thin lines, which is valid for most STIS absorbers).  Dav\'e \&
Tripp obtain $\ggh\sim 10^{-13.3\pm 0.7}\; {\rm s}^{-1}$, with the errors
dominated by systematic uncertainties in modeling, providing among the
most precise constraints on the metagalactic $\ggh$ to date.

\section{Conclusion}

The \lya forest has proved to be a powerful tool for constraining a
wide variety of physical and cosmological parameters, when hydrodynamic
simulations and the FGPA are used in conjunction with high-resolution
quasar spectra.  Much progress has already been made at high redshift,
where the metallicity, ionization state, and mass fluctuation spectrum 
have all been derived from \lya forest data.

Hydrodynamic simulations of structure formation also provide an excellent
description of low redshift \lya forest absorbers, particularly for the
weak absorbers that reside in moderate-overdensity regions analogous
to high-redshift absorers.  HST/STIS has opened up a new era in the
study of the low-redshift IGM, which promises to progress considerably
further with the deployment of the Cosmic Origins Spectrogaph aboard HST
(in 2003).  Initial results indicate that the low-redshift \lya forest
can yield important constraints on physical parameters in the IGM, as
well as provide a more complete picture of the evolution of baryons
in the Universe.


\acknowledgments
I thank Todd Tripp, Lars Hernquist, Neal Katz, David Weinberg, and Ray
Weymann for collaborative efforts on these projects and helpful
discussions.  The Q1422 spectrum is in the top panel.

\end{document}